\documentclass[twocolumn,secnumarabic,amssymb,amsmath,nofootinbib,tightenlines,showpacs,floatfix,superscriptaddress,aps,prl]{revtex4-1}
\usepackage{amsfonts}
\usepackage{color}
\usepackage{txfonts} 
\usepackage{amsbsy}
\usepackage{bm}%
\usepackage[colorlinks=true,linkcolor=blue,filecolor=blue,menucolor=yellow,urlcolor=blue,citecolor=blue,anchorcolor=blue]{hyperref}
\usepackage{graphicx}
\usepackage{dcolumn}
\usepackage{times} 

\begin{document}
\title{Perfect Absorption in Ultrathin Epsilon-Near-Zero Metamaterials Induced by Fast-Wave Non-Radiative Modes}
\author{Simin Feng}
\email{simin.feng@navy.mil}
\author{Klaus Halterman}
\affiliation{Michelson Lab, Physics Division, Naval Air Warfare Center, China Lake, California 93555}
\date{\today}

\begin{abstract}
Above-light-line surface plasmon polaritons can arise 
at the interface between a metal and $\epsilon$-near-zero metamaterial.  
This unique feature induces unusual fast-wave non-radiative modes in a $\epsilon$-near-zero material/metal bilayer.  
Excitation of this peculiar mode leads to wide-angle perfect absorption in low-loss ultrathin metamaterials.  
The ratio of the perfect absorption wavelength to the thickness of the $\epsilon$-near-zero metamaterial can be as high as $10^4$;  
the electromagnetic energy can be confined in a layer as thin as $\lambda/10000$.  Unlike conventional fast-wave leaky modes, 
these fast-wave non-radiative modes have quasi-static capacitive features that 
naturally match with the space-wave field, and thus are easily accessible from free space.  
The perfect absorption wavelength can be tuned from mid- to far-infrared by tuning the $\epsilon\approx0$ wavelength while keeping the thickness of the structure unchanged.
\end{abstract}


\maketitle
A metamaterial is a composite structure with an electromagnetic (EM) response
not readily observed in naturally occurring materials.  Many remarkable phenomenon
have been predicted \cite{Vesalago,Feng,Lai,nguyen}
by tuning the permittivity $\epsilon$ and 
permeability $\mu$ in extraordinary ways.
If the dielectric response is made vanishingly small,
creating an epsilon-near-zero (ENZ) material,
interesting radiative effects are expected to occur \cite{Enoch,Silveirinha,Halterman}. 
On the other hand, by manipulating the EM response to achieve
a small transmittance ($T$) and reflectance ($R$),
enhanced absorption can ensue.
Strong absorption in a thin layer typically requires high loss.  
One of the earliest absorbers, the Salisbury screen \cite{Salisbury}, is
based on the phenomenon of destructive wave interference, and 
is thus limited to a minimum thickness of one quarter wavelength.  
To overcome the thickness constraint, absorbing screens using metamaterials \cite{Engheta}
and high impedance ground planes \cite{Sievenpiper} have recently been proposed.  
By exploring resonant enhancement, thin metamaterial and nanoplasmonic absorbers were demonstrated 
in structures having localized resonances \cite{Landy,XLiu,Brown,Li,Avitzour,Ye,Kazemzadeh,Costa,Kravets,NLiu,Kang,Chern,atwater}.  
In those structures, the geometrical quality factor (GQF), i.e. the ratio 
of the perfect absorption wavelength to the thickness of the medium, is significantly improved compared to the standard Salisbury screen.  
The best GQF of those structures is about 40 \cite{Li}.  

\begin{figure}[htb]
\centering\includegraphics[width=.25\textwidth]{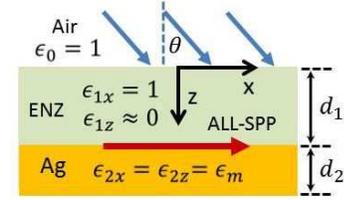}
\caption{(Color online) A plane wave is incident (at an angle $\theta$)
on an anisotropic ($\epsilon_{1x}=1$ and $\epsilon_{1z}\approx0$) ENZ/metal bilayer with thicknesses $d_1$ and $d_2$ for the ENZ medium and metal (Ag), respectively.
Above-light-line surface plasmon polaritons (ALL-SPPs) are excited at the ENZ-metal interface.}
\label{Fig1}
\end{figure}

Based on a fundamentally different mechanism, in this paper we demonstrate wide-angle perfect 
absorption in a low-loss ultrathin ENZ-metamaterial/metal bilayer as shown in Fig.~\ref{Fig1}.  In our structure, the GQF can be as high as $10^4$.  
This structure possesses fast-wave non-radiative (FWNR) modes due to unconventional above-light-line surface plasmon polaritons (ALL-SPPs) at the 
ENZ-metal interface, which is easily accessible from free space.  Fast waves have phase velocities exceeding the speed of light in vacuum.   
Conventional fast waves in planar structures are radiative leaky modes that cannot be excited by plane wave incidence due to wave-vector mismatch 
at the boundary.  For our ENZ-metal structure, the exotic FWNR mode is naturally matched 
to free space at the ENZ-air interface, and 
thus is easily accessible from free space.  Unlike the Salisbury screen where the thickness of the absorbers increases with the absorption 
wavelength, in our structure the wavelength can be tuned from mid- to far-infrared (IR) by tuning the $\epsilon\approx0$ wavelength while 
maintaining the thickness. 

To investigate this absorption phenomenon in a general fashion, 
an anisotropic ENZ metamaterial ($\epsilon_{1x}=1$, $\epsilon_{1z}\approx0$) is assumed (see 
Fig.~\ref{Fig1}).  In the following, the subscripts 1 and 2 refer, respectively, to the ENZ medium and metal.  
The areas above the ENZ material and below the metal are free space and refer to regions 0 and 3, respectively.  
Since the $\epsilon_{1x}$ has no significant impact on the result, $\epsilon_{1x}=1$ is assumed throughout the paper.  
The permittivity of silver in the infrared region is obtained by curve fitting experimental data \cite{Palik} with a Drude model, 
$\epsilon_m=1-\omega_p^2/\omega^2$, where the ``plasma frequency" $\omega_p=5.38\,\mu$m$^{-1}$.
The propagation constant of SPPs at the interface of two semi-infinite anisotropic media is given by
\begin{equation}
\label{Ksp}
\frac{k_{spp}}{k_0} = \sqrt{\dfrac{\epsilon_{1z}\epsilon_{2z} \bigl(\epsilon_{2x}\mu_{1y}-\epsilon_{1x}\mu_{2y}\bigr)} {\epsilon_{2x}\epsilon_{2z}-\epsilon_{1x}\epsilon_{1z}}}
	\approx \sqrt{\epsilon_{1z}\mu_{1y} \left(1-\delta\right)}  \,,
\end{equation}
where $k_0=\omega/c$ and $\delta=\epsilon_{1x}\mu_{2y}/(\epsilon_{2x}\mu_{1y})$.  The approximation is taken when $\epsilon_{1z}\rightarrow0$.  In our case, $\mu_{1y}=\mu_{2y}=1$ and $\epsilon_{2x}=\epsilon_{2z}=\epsilon_m\sim-10^5$ (in the IR region).  Thus, $k_{spp}/k_0\approx \sqrt{\epsilon_{1z}\mu_{1y}}<1$, 
which characterizes above light-line SPP dispersion, and is 
clearly different from conventional SPP dispersion.

The absorption can be calculated from Maxwell's equations.  Assuming a harmonic time dependence $\exp(-i\omega t)$ for the EM field, we have
\begin{eqnarray}
\label{Maxw}
\begin{split}
\nabla\times\bigl( \bar{\bar\mu}_n^{-1} \cdot \nabla\times{\bm E}\bigr)  &=\, k_0^2\bigl(\bar{\bar\epsilon}_n \cdot{\bm E}\bigr)  \,,\\
\nabla\times\bigl( \bar{\bar\epsilon}_n^{-1} \cdot \nabla\times{\bm H}\bigr)  &=\, k_0^2\bigl(\bar{\bar\mu}_n \cdot{\bm H}\bigr)  \,,
\end{split}
\end{eqnarray}
where $\bar{\bar\epsilon}_n$ and $\bar{\bar\mu}_n$ are, respectively, the permittivity and permeability 
tensors for a given (uniform) region ($n=0,1,2,\cdots$), which in the principal coordinates are described by,
\begin{eqnarray}
\bar{\bar\epsilon}_n &=& \epsilon_{nx}\hat{\bm x}\hat{\bm x} + \epsilon_{ny}\hat{\bm y}\hat{\bm y} + \epsilon_{nz}\hat{\bm z}\hat{\bm z}  \,,\cr
\bar{\bar\mu}_n &=& \mu_{nx}\hat{\bm x}\hat{\bm x} + \mu_{ny}\hat{\bm y}\hat{\bm y} + \mu_{nz}\hat{\bm z}\hat{\bm z}  \,.
\end{eqnarray}
We consider TM modes, corresponding to non-zero field components $H_y$, $E_x$, and $E_z$.  The magnetic field $H_y$ satisfies
the following wave equation:
\begin{equation}
\frac{1}{\epsilon_z} \frac{\partial^2H_y}{\partial x^2} + \frac{1}{\epsilon_x} \frac{\partial^2H_y}{\partial z^2} + k_0^2\mu_yH_y = 0  \,,
\end{equation}
which admits solutions of the form $\psi(z)\exp(i\beta x)$.  
The parallel wave-vector $\beta$ is determined by the incident wave, and is conserved across the interface,
\begin{equation}
\label{beta}
\beta^2 = k_0^2\epsilon_{nz}\mu_{ny} - \alpha_n^2 \frac{\epsilon_{nz}}{\epsilon_{nx}}  \,,\hskip.2in  (n=0,1,2,\cdots)  \,,
\end{equation}
where $\alpha_n$ is the wave number in the $z$ direction.  The functional form of $\psi(z)$ is either a simple exponential $\exp(i\alpha_nz)$ for the semi-infinite region or a superposition of $\cos(\alpha_nz)$ and $\sin(\alpha_nz)$ terms for the bounded region (along the $z$ direction).  The other two components $E_x$ and $E_z$ are found using Maxwell's equations.  By matching boundary conditions at the interface, i.e., continuity of $H_y$ and $E_x$, the transmittance ($T$) and reflectance ($R$) can be calculated via the Poynting vector $\bm{S}$, given by $\bm{S}=c/8\pi\Re(\bm{E}\times\bm{H}^*)$.  
The fraction of energy that is absorbed by the system is determined by the absorptance ($A$), where $A=1-T-R$,  consistent with energy conservation.
\begin{figure}[htb]
\centering\includegraphics[width=.38\textwidth]{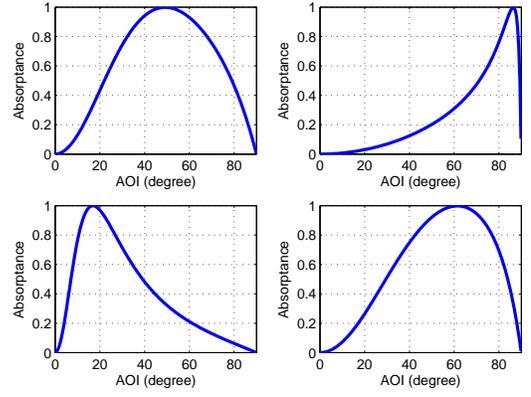}
\caption{(Color online) Absorptance vs. angle of incidence (see Fig.~\ref{Fig1}) at the wavelength $\lambda=10.94\,\mu$m (left panels) and $\lambda=201.66\,\mu$m (right panels) for
two different  thicknesses of the ENZ medium: $d_1=0.02\,\mu$m (top panels) and $d_1=0.2\,\mu$m (bottom panels).  The Ag-layer thickness ($d_2$) is $20\,$nm.}
\label{Fig2WBabs6}
\end{figure}
Figure~\ref{Fig2WBabs6} shows the absorptance of the ENZ-metal structure versus the angle of incidence (AOI) in the mid- to far-IR regime. Unless
stated otherwise, the results that follow correspond to $\epsilon_{1z}=0.001+i0.01$.
The AOI at which the perfect absorption occurs depends on the wavelength and the thickness of the ENZ layer.  
The full-width-half-maximum (FWHM) angular width can be as high as $\sim60^\circ$ (top-left panel), 
while the GQF can reach $\sim10^4$ (top-right panel).  
When the thickness increases, the angular bandwidth varies differently in the mid- and far-IR regions.
\begin{figure}[htb]
\centering\includegraphics[width=.38\textwidth]{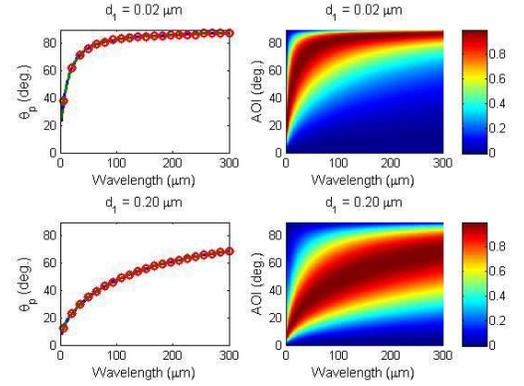}
\caption{(Color online) Absorptance (right panels) and direction of perfect absorption (left panels) when the ENZ-thickness $d_1=0.02\,\mu$m (top panels) and $d_1=0.2\,\mu$m (bottom panels).  Color-bars represent the magnitude of the absorptance.  The curves in the left panels are computed using three different methods (see text).  
Solid (blue): numerically extracted from the corresponding 2D plot in the right panels.  Dashed (green): obtained from Eq.~(\ref{Zeff}).  
Circles (red): computed from Eq.~(\ref{Root}).
The Ag thickness is $20\,$nm.}
\label{Fig3WBabs8}
\end{figure}

In Fig.~\ref{Fig3WBabs8}, we show how the absorptance varies with the AOI and incident wavelength (right panels).
The left set of panels corresponds to the incident angle $\theta_p$ that results in perfect absorption, as a function
of wavelength.  Perfect absorption occurs when the $\theta$ dependent effective impedance (${\cal Z}_e$) of the ENZ-metal structure matches that of free space (${\cal Z}_0$), 
i.e. ${\cal Z}_e={\cal Z}_0$, in which case the reflection coefficient is zero.  The effective impedance of the ENZ-metal structure can be expressed as
\begin{equation}
\label{Zeff}
{\cal Z}_e = {\cal Z}_1 \frac{1-r_{12}\exp\bigl(i2\phi\bigr)} {1+r_{12}\exp\bigl(i2\phi\bigr)} = {\cal Z}_0  \,,
\end{equation}
where $r_{12}=({\cal Z}_1-{\cal Z}_2)/({\cal Z}_1+{\cal Z}_2)$, ${\cal Z}_j=\alpha_j/\epsilon_{jx}\,(j=0,1,2)$, and  $\phi=\alpha_1d_1$.
Perfect absorption occurs at the particular AOI that satisfy Eq.~(\ref{Zeff}).  
It is also informative to analyze the corresponding waveguide modes, which are solutions to the transcendental equation,
\begin{equation}
\label{Root}
\tan\phi = -i\frac{{\cal Z}_1 ({\cal Z}_0+{\cal Z}_2)} {{\cal Z}_1^2+{\cal Z}_0{\cal Z}_2},
\end{equation}
which implicitly depends on the parallel complex propagation constant $\beta$ (Eq.~(\ref{beta})).  The mode solutions of Eq.~(\ref{Root}) have four 
branches due to the $\pm$ signs inherent to the square root of $\alpha_0$ and $\alpha_2$ in Eq.~(\ref{beta}).  
The branch leading to perfect absorption corresponds to both $\alpha_0<0$ and $\alpha_2<0$.  
In the long wavelength regime and for ultrathin slabs, this branch provides fast-wave ($\beta/k_0<1$), non-radiative ($\beta$ real) modes that
are guided along the ENZ layer.  
In the left panels of Fig.~\ref{Fig3WBabs8} we compute the direction of perfect absorption as a function of wavelength in three complimentary ways:
The solid (blue) curves are numerically extracted from the absorptance corresponding to the right panels, 
the dashed (green) curves are obtained from Eq.~(\ref{Zeff}), where the structure is impedance matched to free space,
and the circles (red) are calculated using Eq.~(\ref{Root}) and a root searching algorithm. 
The overlap of each set of data further demonstrates that the underlying mechanism of perfect absorption is the excitation of FWNR modes.  
We see also that perfect absorption occurs only at oblique incidence where 
the electric field has a nonzero component in the normal direction.

The energy density, $U$, in lossy and dispersive anisotropic media is given by,
\begin{equation} 
\label{U}
U = \frac{1}{16\pi} \Re\left[{\bm E}^\dagger\cdot\frac{\partial(\omega\bar{\bar\epsilon})}{\partial\omega}{\bm E} +  {\bm H}^\dagger\cdot\frac{\partial(\omega\bar{\bar\mu})}{\partial\omega}{\bm H}\right]. 
\end{equation}
\begin{figure}[htb]
\centering\includegraphics[width=.38\textwidth]{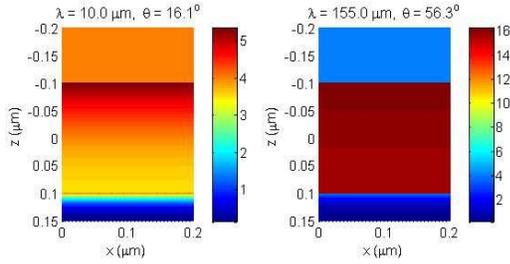}
\caption{(Color online) Spatial distribution of the energy density computed from Eq.~(\ref{U}) (normalized so that $H_{y}$ is unity at the ENZ-air interface)
for $\lambda=10\,\mu$m and AOI = $16.1^\circ$ (left) and $\lambda=155\,\mu$m and AOI = $56.3^\circ$ (right).  The ENZ-layer with $d_1=0.2\,\mu$m is located at $z\in[-0.1\ 0.1]$.  
The regions $z\in[0.1\ 0.15]$ and $z\in[-0.2\ -0.1]$ are, respectively, silver and air.  Color-bars represent the magnitude ($\times0.01$) of the energy density, 
indicating the energy is more confined to the ENZ medium at far-IR wavelengths.}
\label{Fig4WBabs9}
\end{figure}
\begin{figure}[htb]
\centering\includegraphics[width=.38\textwidth]{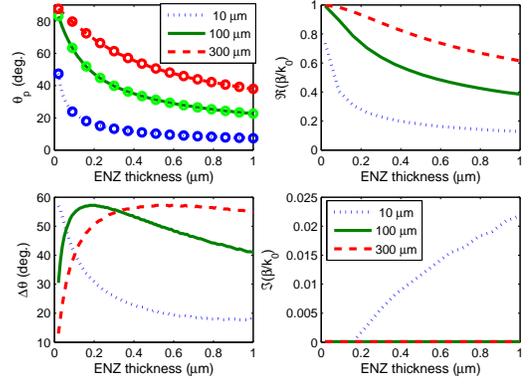}
\caption{(Color online) Direction (top left) and FWHM bandwidth (bottom left) of  perfect absorption  vs. ENZ thickness.  
The curves are numerically extracted from the absorptance when $\lambda=10\,\mu$m (dots (blue)), $\lambda=100\,\mu$m (solid (green)), 
and $\lambda=300\,\mu$m (dashed (red)). The circles in the left-top panel are obtained from Eq.~(\ref{Root}).  
The normalized propagation constant $\beta/k_0$ corresponding to these modes is shown in the right panels.  
Top-right: real part of $\beta/k_0$.  Bottom-right: imaginary part of $\beta/k_0$.  The Ag thickness ($d_2$) is $10\,$nm.}
\label{Fig5WBthk5}
\end{figure}
We show in Fig.~\ref{Fig4WBabs9} the spatial distribution of the energy density
in the ENZ-metal structure under conditions of perfect absorption at mid- and far-IR wavelengths.  
In the mid-IR (left panel), the field inside the ENZ-layer is concentrated at the two interfaces, and
a very thin layer of ALL-SPPs can be seen at the ENZ-metal interface.  The strong field 
at the ENZ-air interface is distinguishable from conventional surface waves, since in this case  
the exponential decay of the field occurs strictly inside the ENZ medium.  Outside of the ENZ medium, 
the field neither decays like a surface wave nor does it increase like a leaky wave.  
The observed uniformity indicates that the field outside the structure 
is plane wave matching with the free-space wave, and thus can be easily excited by plane wave incidence -- a 
striking difference from surface waves and leaky waves.  In the far-IR (right panel), the field is even more uniform 
and confined to the ENZ-medium and also matched to the space-wave field at the ENZ-air boundary.  Around the 
point where the real part of $\epsilon$ approaches zero, even with a small loss $\Im(\epsilon_{1z})=0.01$, 
the presence of the FWNR mode reinforces the increasing absorption due to the strong field inside 
the ENZ medium (by virtue of the electric displacement continuity) 
leading to perfect absorption in the ultrathin ENZ medium.

The minimum thickness of ENZ materials is generally limited by fabrication methods except in some cases of naturally occurring resonances.  
Figure~\ref{Fig5WBthk5} shows the direction, $\theta_p$ (top left), and FWHM angular bandwidth (bottom left) of perfect 
absorption  versus ENZ thickness for three wavelengths, as well as the corresponding normalized complex propagation constants (right panels).  
The circles in the top left panel represent solutions to Eq.~(\ref{Root}), 
which are consistent with the absorption calculated using the Poynting vector.  
These curves also correlate with the solutions from the matched impedance  expression (Eq.~(\ref{Zeff})) (not shown).  
When the thickness increases, the direction of perfect absorption is shifted towards the surface normal for both mid- and far-IR regions. 
There is an optimal thickness that yields the largest bandwidth.  
The real part of $\beta/k_0$ is above the vacuum light-line, indicating a fast-wave character for all three 
wavelengths shown.  The imaginary component of $\beta$ is zero for the far-IR wavelengths.  
For $\lambda=10\,\mu$m, $\Im(\beta)=0$ when $d_1<0.2\,\mu$m.  
When $d_1>0.2\,\mu$m, $\beta$ starts to pick up a small imaginary part which increases with thickness:
The fast-wave non-radiative mode transforms into a fast-wave leaky mode \cite{Halterman} as the thickness increases. 
\begin{figure}[htb]
\centering\includegraphics[width=.45\textwidth]{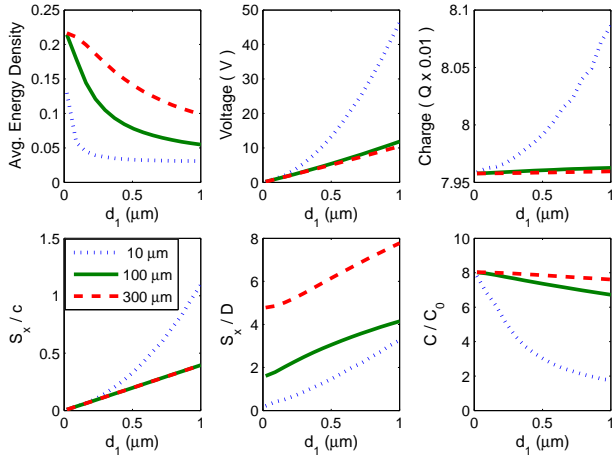}
\caption{(Color online) Average energy density (left top), transport energy ($S_x/c$) per unit area (left bottom), voltage ($V$) across the ENZ-layer (middle top), the ratio of power to dissipation rate (middle bottom), 
the effective surface charge density $Q$ (right top) and the normalized capacitance 
$C/C_0$ (right bottom) vs. $d_1$ for $\lambda=10\,\mu$m (dots (blue)), $\lambda=100\,\mu$m (solid (green)), and $\lambda=300\,\mu$m (dashed (red)).}
\label{Fig6WBthk7}
\end{figure}
\begin{figure}[htb]
\centering\includegraphics[width=.39\textwidth]{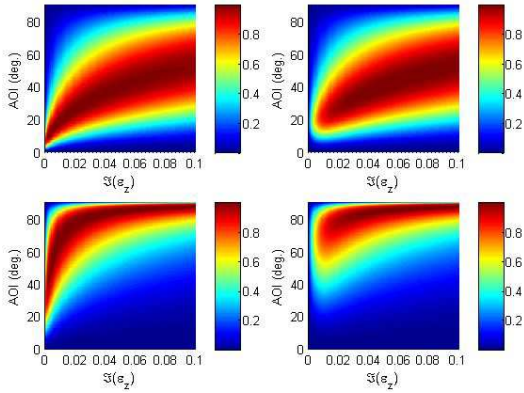}
\caption{(Color online) Color map of the absorptance versus AOI and loss ($\Im(\epsilon_{1z})$) 
for $\lambda=10\,\mu$m (top panels) and $\lambda=200\,\mu$m (bottom panels). The left and right
set of  panels correspond to 
$\Re(\epsilon_{1z})=0.001$ and $\Re(\epsilon_{1z})=0.01$ respectively.  
Here, $d_1=0.15\,\mu$m and $d_2=10\,$nm.  The color bars represent the magnitude of the absorptance.}
\label{Fig7WBepz2}
\end{figure}

To further understand the FWNR modes, we integrated the transport power $S_x$, dissipation rate $D$, and $\Re(E_z)$ 
over the ENZ-thickness (normalized by the input $E_z$).  The dissipation rate is given by, 
\begin{equation} 
\label{D}
D = \frac{\omega}{8\pi} \Im\Bigl[{\bm E}^\dagger\cdot \bar{\bar\epsilon} {\bm E} + {\bm H}^\dagger\cdot \bar{\bar\mu} {\bm H}\Bigr].
\end{equation}
The average energy density in the ENZ medium is also calculated (Eq.~(\ref{U})), as well as the effective surface charge density $Q=\epsilon_{1z} E_z/4\pi$ (normalized by the input $E_z$) and capacitance $C=Q/V$ per unit area, 
which is much larger than the capacitance ($C_0=\epsilon_{1z}/d_1$) of a parallel-plate capacitor (of unit-area), due in part to the field enhancement inside the ENZ 
material.  These results are presented in Fig.~\ref{Fig6WBthk7}.  The confinement is stronger for far-IR and decreases with the thickness (left-top).  The linear responses of the voltage (middle-top) 
and constant behavior of the charge (right-top) as a function of thickness are hallmarks of capacitors.  
The deviation from the linear behavior for $\lambda=10\,\mu$m can be understood as a balance between radiative loss, which 
acquires more ``charge'' stored in the ``capacitor'' (right- and middle-top) and energy transport (left-bottom) (see right-bottom panel of Fig.~\ref{Fig5WBthk5} where 
$\Im(\beta)\ne0$ when $d_1>0.2\,\mu$m for $\lambda=10\,\mu$m).  
The radiative loss reduces the enhancement of the capacitance at $\lambda=10\,\mu$m as shown in the right-bottom panel. 
In essence, the higher the confinement, the larger the capacitance.  
The effective capacitance is found to be equal to $cQ^2/(2S_x)$ or $2S_x/(cV^2)$, which implies the transport energy is driven by quasi-static  ``surface charges''.  
Thus, FWNR modes can be considered quasi-static, capacitively driven fast waves.  
The middle-bottom panel of Fig.~\ref{Fig6WBthk7} 
shows the ratio of transport power to dissipation power, suggesting longer wavelengths have better power handling capability 
if the structure is used as an ultrathin channel to transport light.  The influence of loss on the absorption is illustrated 
in Fig.~\ref{Fig7WBepz2} for different $\epsilon_{1z}$ (real part)  in the mid- and far-IR.  In the far-IR, the smaller 
$\Im(\epsilon_{1z})$ has larger angular bandwidth, which is opposite to what occurs in the mid-IR.

In summary, wide-angle perfect absorbers have been demonstrated in ultrathin ENZ-metal structures, and which have extraordinary high 
geometric quality factors.  Such structures may have applications involving ultra light-weight 
infrared absorbers, cloaking, EM shielding, photovoltaic systems, and highly
efficient infrared sensors and detectors.  The authors gratefully acknowledge the sponsorship of ONR N-STAR and NAVAIR's ILIR programs.

\end{document}